\documentclass[runningheads]{llncs}
\usepackage[T1]{fontenc}
\usepackage{graphicx}
\usepackage{tabularx}
\usepackage{booktabs}
\usepackage{amsmath}
\usepackage{amssymb}
\usepackage{soul}
\usepackage{xcolor}
\soulregister\cite7
\soulregister\ref7
\begin{document}
\title{Improving WiFi CSI Fingerprinting with IQ Samples}
\author{Junjie Wang\inst{1} \and
Yong Huang\inst{1} \thanks{*Corresponding author} \and
%Yong Huang\inst{1} \and
Feiyang Zhao\inst{1} \and
Wenjing Wang\inst{1} \and 
Dalong Zhang\inst{1} \and
Wei Wang\inst{2}}

%Third Author\inst{3}\orcidID{2222--3333-4444-5555}
\authorrunning{J. Wang et al.}
\institute{Zhengzhou University, Zhengzhou, China\\
\email{yonghuang@zzu.edu.cn} \and
Huazhong University of Science and Technology, Wuhan, China}
\maketitle              % typeset the header of the contribution
\begin{abstract}
Identity authentication is crucial for ensuring the information security of wireless communication. 
Radio frequency (RF) fingerprinting techniques provide a prom-ising supplement to cryptography-based authentication approaches but rely on dedicated equipment to capture in-phase and quadrature (IQ) samples, hindering their wide adoption.
Recent advances advocate easily obtainable channel state in-formation (CSI) by commercial WiFi devices for lightweight RF fingerprinting, but they mainly focus on eliminating channel interference and cannot address the challenges of coarse granularity and information loss of CSI measurements. 
To overcome these challenges, we propose CSI\textsuperscript{2}Q, a novel CSI fingerprinting sys-tem that achieves comparable performance to IQ-based approaches. 
Instead of ex-tracting fingerprints directly from raw CSI measurements, CSI\textsuperscript{2}Q first transforms them into time-domain signals that share the same feature space with IQ samples. 
Then, the distinct advantages of an IQ fingerprinting model in feature extraction are transferred to its CSI counterpart via an auxiliary training strategy. 
Finally, the trained CSI fingerprinting model is used to decide which device the sample under test comes from. We evaluate CSI\textsuperscript{2}Q on both synthetic and real CSI datasets. 
On the synthetic dataset, our system can improve the recognition accuracy from 76\% to 91\%. On the real dataset, CSI\textsuperscript{2}Q boosts the accuracy from 67\% to 82\%.

\end{abstract}
\section{Introduction}
Nowadays, WiFi has become one of the most essential communication technologies that connect various wireless devices, such as smartphones, tablets, and smart speakers.
Although WiFi has been indispensable in many aspects of our daily lives, it is susceptible to identity-based attacks like spoofing and impersonating, thus threatening the information security of communication system.
The IEEE 802.11 protocol has provided cryptography-based authentication schemes, but their effectiveness has proven to be inadequate~\cite{36,37}.
In recent years, radio frequency (RF) fingerprinting has been proposed as an important supplement to cryptography-based security technologies in wireless networks~\cite{4}.
%In any wireless communication system, baseband signals are processed by transmitter components such as oscillators and power amplifiers to become radio frequency (RF) signals.
%Hardwares involved in this process inevitably have imperfections, including carrier frequency offset in oscillators and nonlinearity in power amplifiers.
%These hardware impairments are collectively referred to as RF fingerprints.
%RF fingerprints are device-specific, difficult to forge and tamper with \cite{13}, thus being used as features for device recognition.
%Furthermore, compared to traditional cryptography-based wireless network security technologies, RF fingerprinting technology has lower algorithm complexity and does not require additional key management.
It utilizes hardware imperfections, such as carrier frequency offset in oscillators and nonlinearity in power amplifiers, to identify mobile devices.
These hardware impairments are referred to as RF fingerprints, which are device-specific and difficult to forge~\cite{13}.
Additionally, compared to cryptography-based approaches, RF fingerprinting has lower requirements on computational complexity and communication overhead~\cite{6}, making it suitable for many miniaturized WiFi devices.

Currently, RF fingerprinting solutions rely on time-domain in-phase and quadrature (IQ) samples.
Handcrafted features are extracted from IQ signals as RF fingerprints for device identification~\cite{15,27}.
Additionally, raw IQ signals are directly fed into deep neural networks for automatic RF fingerprint extraction~\cite{9,18}.
While IQ-based solutions generally demonstrate high identification performance, their practical application in the real world is impeded by their requirement of costly and dedicated RF equipment like universal software radio peripherals (USRPs).

Growing attempts have been devoted to exploring RF fingerprinting based on channel state information (CSI) that is easily obtainable for commercial WiFi devices~\cite{30,31}.
A CSI measurement mainly describes how a WiFi signal propagates from a transmitter to a receiver, but it also carries RF fingerprints due to hardware imperfections. 
%In recent years, several CSI collection tools~\cite{29, 30, 31} have been developed for wireless sensing and localization, making CSI measurements easily obtainable for commercial WiFi devices.
Existing CSI fingerprinting schemes mainly focus on mitigating channel interference.
Some studies demonstrate that the channel-independent features such as the carrier frequency offset~\cite{21}, phase error~\cite{22}, and power variance caused by the power amplifier~\cite{7} exist in CSI measurements.
The literature~\cite{24} realizes CSI-based RF fingerprinting by eliminating channel interference.
However, besides channel interference, the task of WiFi CSI fingerprinting still faces another two challenges, and none of the existing works have addressed them.
\begin{itemize}
    \item \textbf{Coarse Granularity.}
    According to the IEEE 802.11 protocol, a WiFi packet contains three parts, i.e., Preamble, SIGNAL, and DATA, in the time domain.
    Only IQ samples in the preamble are used for channel estimation.
    Even worse, one CSI measurement contains the channel responses of specified subcarriers.
    For example, for a 2.4~GHz WiFi with a 20~MHz bandwidth, the preamble signal consists of 320 IQ samples but a CSI measurement has only 52 channel estimates of subcarriers.
    Thus, the coarse-grained feature representation renders it harder to extract subtle RF fingerprints from CSI measurements. 
    \item \textbf{Information Loss.}
    CSI measurements are the outcome of a series of processes, such as fast Fourier transform (FFT), on preamble IQ samples.
    Throughout these processes, deviations are inevitably introduced in CSI measurements, thus blurring the hardware imperfection effect and resulting in the loss of device fingerprint information. 
\end{itemize}

In this paper, we propose CSI\textsuperscript{2}Q, a novel CSI fingerprinting system that achieves comparable performance to IQ-based approaches. Besides channel interference, CSI\textsuperscript{2}Q tackles the other two challenges at the same time. 
The core idea of CSI\textsuperscript{2}Q is to transform frequency-domain CSI measurements into time-domain samples and transfer knowledge about RF fingerprints from an IQ-based model to its CSI-based counterpart via deep auxiliary learning.
To achieve this goal, we propose three effective components in our system.
\textit{First}, since the time duration of a WiFi packet is typically within the channel coherence time, the channel responses of all subcarrier signals are correlated, especially for adjacent subcarriers.
Thus, we propose a cyclic shift division scheme on all estimates in one CSI measurement for mitigating channel interference.
\textit{Second}, motivated by the relation between received preamble signals and true channel responses, we transform the processed CSI data into time-domain signals that share the same feature space as preamble IQ samples.
\textit{Lastly}, we introduce an auxiliary learning approach, where the distinct advantages of an IQ fingerprinting model in feature extraction are transferred to its CSI counterpart.
By doing so, the CSI fingerprinting model can capture fine-grained features that could be missed by it alone.
We build an auxiliary IQ dataset and a synthetic CSI dataset involving 85 wireless devices based on the public WiSig dataset~\cite{11}. 
%We evaluate CSI\textsuperscript{2}Q on the synthetic CSI dataset.
%Our system can improve the recognition accuracy from 76\% to 91\%.
Moreover, we utilize a laptop running PicoScenes~\cite{31} to collect CSI measurements from ten commercial WiFi routers and obtain a real CSI dataset. 
We evaluate CSI\textsuperscript{2}Q on both the synthetic and real CSI datasets.
On the synthetic dataset, our system can improve the recognition accuracy from 76\% to 91\%.
On the real dataset, CSI\textsuperscript{2}Q boosts the average accuracy from 67\% to 82\%.

The main contributions of this work are summarized as follows.
\begin{itemize}
    \item We show that CSI and IQ data are inherently correlated, and CSI fingerprinting can benefit from IQ samples for performance improvement.
    \item We theoretically analyze the reason why the original CSI measurements are not as effective as the time domain IQ samples in RF fingerprinting. And we propose CSI\textsuperscript{2}Q that addresses the challenges of channel interference, coarse granularity and information loss in CSI fingerprinting, achieving comparable device identification performance to IQ-based approaches.
    \item We evaluate CSI\textsuperscript{2}Q on both the synthetic and real CSI datasets.
    The experimental results show that our system can effectively improve the recognition performance of CSI fingerprinting models.
\end{itemize}

\section{Threat Model and CSI Features}
\subsection{Threat Model}
We consider a common scenario where a WiFi access point (AP) is deployed to provide Internet access for a set of wireless devices, such as smart light bulbs, air conditioners, and so on.
Each device has a unique identity and is assigned a security key for successful association and authentication. 
In this scenario, we consider identity-based attacks, where a malicious device could impersonate another one by forging its identity, such as MAC and IP addresses, and stealing its security key.
After sneaking into the WiFi network, it will either inject fake data into a remote server or load the valuable files from it.

To defend against such attacks, we consider physical layer authentication based on radio frequency fingerprints.
First, we collect $N$ CSI measurements $\mathcal{H}=  \left\{H_n\right\}^{N}_{n=1} $ from $I$ registered devices $\mathcal{D} = \left\{D_i \right\}^{I}_{i=1} $.
Therein, $D_i\in  \left\{0,1\right\}^{I} $ is a one-hot vector indicating the device index.
Then, a classification model is trained based on the dataset $\mathcal{H}$.
Next, given a new CSI measurement $H$, the model recognizes from which device the measurement comes.

\subsection{CSI Features}
%Differing from RF fingerprinting schemes that rely on IQ signals, we attempt to leverage channel state information (CSI) to identify wireless devices.
%In this subsection, we introduce the basics of CSI features and conduct a preliminary study to show the feasibility and difficulty of WiFi CSI fingerprinting.

%\subsubsection{CSI Basics.} 
Channel state information (CSI) is utilized to characterize the propagation characteristics of 
wireless signals, reflecting their amplitude attenuation, phase variation, and time delay from the transmitter to the receiver.
Typically, the orthogonal frequency division multiplexing (OFDM) technique is adopted to divide the WiFi band into $K$ orthogonal subcarriers for high throughput.
In this condition, CSI is represented in a matrix form in the frequency domain.
Given $k$-th subcarrier, let $x_k$ be the transmitted signal vector, and $y_k$ be the received signal vector. 
Moreover, $F(\cdot)$ indicates the impact of the transmitter's hardware imperfection, $\sigma$ represents the noise vector, and $h_{c_k}$ is the true channel response.
The relationship between $x_k$ and $y_k$ can be expressed as
\begin{equation}
  y_k = h_{c_k} \cdot F(x_k) + \sigma.
\end{equation}
The CSI of $k$-th subcarrier, denoted as $h_k$, is estimated by $h_k = \frac{y_k}{x_k}$.
Therefore, $h_k$ can be approximated as
\begin{align}
  h_k = {\frac{h_{c_k} \cdot F(x_k) + \sigma}{x_k}} \approx h_{c_k} \cdot {G(x_k)}, \label{H_k}
\end{align}
where $G(x_k) = \frac{F(x_k)}{x_k}$ includes the impact of the transmitter’s RF fingerprints. 
The above analysis shows that $h_k$ is an approximation of the channel response $h_{c_k}$. 
Like IQ data, CSI data is also affected by the transmitter’s hardware imperfection, making it feasible to perform RF fingerprinting using CSI measurements.

%\subsubsection{Preliminary Study.}
%\begin{figure}[t]
%  \centering
%  \includegraphics[width=\textwidth]{feature_visualization_0331.pdf}
%  \caption{Visualization of high-dimensional features from IQ and CSI data. The same-color samples belong to one device.}
%  \label{fig:feature visualization}
%\end{figure}
%We conduct some preliminary experiments to demonstrate the feasibility of CSI-based device classification.
%To achieve this, we randomly take 3000 IQ samples of ten devices from WiSig~\cite{11}, a recently released IQ dataset, and further calculate corresponding CSI measurements using the standard channel estimation process.
%Then, we feed the two types of data into the same neural network, respectively, and exploit the t-distributed Stochastic Neighbor Embedding (t-SNE) algorithm to visualize their high-dimensional features extracted by the network.
%As shown in Figure~\ref{fig:feature visualization}, the IQ data of the same device yield feature samples that are more tightly clustered.
%Meanwhile, the boundaries between different clusters are more pronounced, making them easier to distinguish.
%Although the feature samples from the CSI data also exhibit clustering patterns, the CSI data of the same device are more dispersed and stay closer to that of other devices.
%The above observation suggests that RF fingerprints are more difficult to extract from CSI features than IQ signals. 
%In conclusion, CSI-based RF fingerprinting is feasible but more challenging.
\section{System Design}

\subsection{System Overview}
\begin{figure}[t]
  \centering
  \includegraphics[width=\textwidth]{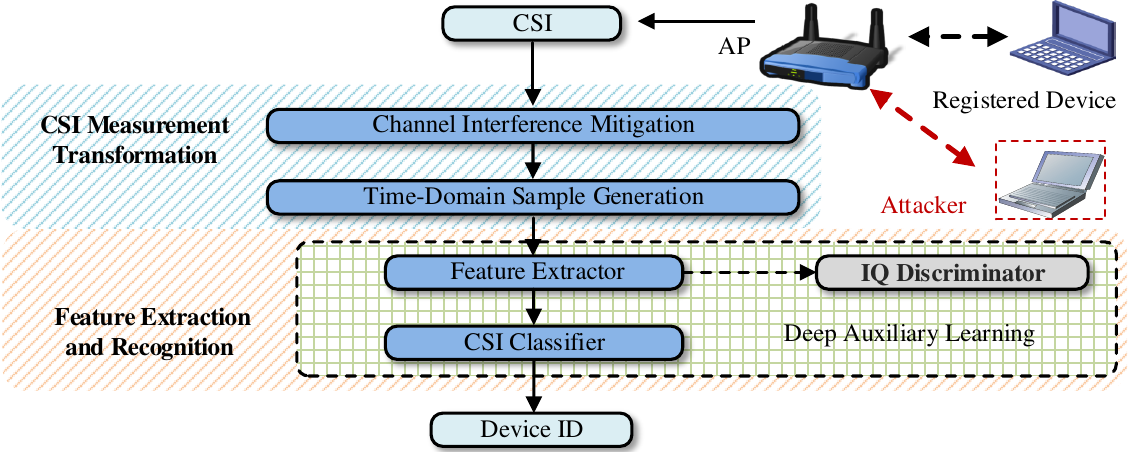}
  \caption{Workflow of CSI\textsuperscript{2}Q. The dashed gray component exists only in the training phase.}
  \label{fig:system_design}
\end{figure}

CSI\textsuperscript{2}Q is a novel CSI-based RF fingerprinting system that achieves comparable performance to IQ-based approaches. 
The core idea of CSI\textsuperscript{2}Q is to transform frequency-domain CSI measurements into time-domain samples and transfer knowledge about RF fingerprints from an IQ-based model to its CSI-based counterpart via deep auxiliary learning.
To realize this goal, the workflow of CSI\textsuperscript{2}Q consists of two phases, i.e., the training phase and the inference phase.
In the training phase, a CSI dataset from registered devices and an auxiliary IQ dataset are leveraged to collaboratively train our fingerprinting model.
In the inference phase, given a new CSI sample, the trained model recognizes from which device the CSI sample comes. 
As illustrated in Figure~\ref{fig:system_design}, the proposed system consists of two main components: CSI Measurement Transformation and Feature Extraction and Recognition.
\begin{itemize}
   \item \textbf{CSI Measurement Transformation.}
   First, we perform cyclic shift division on one CSI measurement for channel interference mitigation. 
   Then, we combine the processed CSI measurement with the standard short and long training sequences to generate a high-dimensional time-domain feature vector that shares the same feature space as the preamble IQ samples.
   \item \textbf{Feature Extraction and Recognition.}
   Based on the generated feature vector, we devise a dual-task learning model with a feature extractor, a CSI classifier, and an IQ discriminator for effective feature extraction.
   In the training phase, an auxiliary learning approach is leveraged to improve the feature extraction ability of our CSI classifier.
   In the testing phase, we only put CSI measurements into the trained classifier for recognizing devices.
\end{itemize}
\subsection{CSI Measurement Transformation}

\subsubsection{Channel Interference Mitigation.}
As aforementioned, a CSI measurement mainly reflects the characteristics of the propagating channel, and thus underlying transmitter RF fingerprints are highly disturbed by them. 
To deal with this issue, the first step of CSI\textsuperscript{2}Q is to mitigate channel interference in a CSI measurement. 
For WiFi with a bandwidth of 20~MHz, the number of orthogonal subcarriers is 52.
In this setting, a CSI measurement between one transmitting antenna and one receiving antenna consists of 52 complex values, each of which corresponds to a frequency response of one subcarrier.
More formally, let $H \in \mathbb{C}^{52}$ be one CSI measurement, which can be denoted as
\begin{equation}
  H = [h_1,\ h_2,\ \dots,\ h_{52}].
\end{equation}
According to Equation~\eqref{H_k}, one estimated channel response of $H$ can be expressed by $h_k = h_{c_k} \cdot {G(x_k)}$,
where $k=1,\ 2,\ \dots,\ 52$.
Since the time duration of one WiFi packet falls within the channel coherence time, it can be assumed that the channel responses experienced by all subcarrier signals are highly correlated, especially for those of adjacent subcarriers.
If we divide one estimated channel response with another, the impact of channel characteristics is likely to be alleviated.
Thus, we use a cyclic shift division scheme for mitigating channel interference.
In this scheme, when $k= 2,\ \dots,\ 52$, we divide $h_k$ with $h_{k-1}$ as
\begin{align}
  \widetilde{h_k} = \frac{h_k}{h_{k-1}} = \frac{{h_{c_k}} \cdot {G(x_k)}}{{h_{c_{k-1}}} \cdot {G(x_{k-1})} } \approx \frac{G(x_k)}{G(x_{k-1})}.
\end{align}
When $k=1$, we have $\widetilde{h_1} = \frac{h_1}{h_{52}} \approx \frac{G(x_1)}{G(x_{52})}$.
So we can observe that the impact of channel characteristics could be removed, and the transmitter's RF fingerprints remain in $\widetilde{h_k}$.
A processed CSI measurement $\widetilde{H} \in \mathbb{C}^{52} $ can be obtained as
\begin{equation}
  \widetilde{H}=[\widetilde{h}_1,\ \widetilde{h}_2,\ \dots,\ \widetilde{h}_{52}]. 
\end{equation}

\subsubsection{Time-Domain Sample Generation.} 
After channel interference mitigation, we proceed to transform a processed CSI measurement into time-domain features like IQ data.  
According to the IEEE 802.11 protocol, the physical layer convergence procedure (PLCP) preamble field is used for channel estimation.
For a 20 MHz bandwidth WiFi signal, the duration of a PLCP preamble field is 16~\textmu s, with short training sequence (STS) and long training sequence (LTS) lasting 8~\textmu s, respectively.
In this condition, the preamble consists of a total of 320 time-domain IQ points, which are further utilized to yield a CSI measurement with 52 complex values.
Compared to corresponding IQ data, a CSI measurement is coarser-grained, rendering it more difficult to extract underlying RF fingerprints. 
To tackle the challenge of coarse granularity, we utilize processed CSI data to generate time-domain signals that have the same dimension as preamble IQ samples.

First, we deduce the relation between IQ samples and true channel responses.  
According to the IEEE 802.11 specification, a short OFDM training symbol $S \in \mathbb{C}^{52}$ consists of 12 effective subcarriers for 20~MHz channel spacing.
In the frequency domain, it can be represented as
\begin{equation}
  \begin{aligned}
  S =& {\sqrt{(13/6)}} \times \left\{0,\ 0,\ 1+j,\ 0,\ 0,\ 0,\ -1-j,\ 0,\ 0,\ 0,\ 1+j,\right.\\ 
  &\left. 0,\ 0,\ 0,\ -1-j,\ 0,\ 0,\ 0,\ -1-j,\ 0,\ 0,\ 0,\ 1+j,\ 0,\ 0,\ 0,\right.\\ 
  &\left. 0,\ 0,\ 0,\ -1-j,\ 0,\ 0,\ 0,\ -1-j,\ 0,\ 0,\ 0,\ 1+j,\ 0,\ 0,\ 0,\right.\\ 
  &\left.  1+j,\ 0,\ 0,\ 0,\ 1+j,\ 0,\ 0,\ 0,\ 1+j,\ 0,\ 0\right\},
    \end{aligned}
\end{equation}
where $\sqrt{(13/6)}$ is a power normalization factor.
In the time domain, the standard STS signal $s_S(t)$ can be generated using the following formula as
\begin{equation}
  s_S(t) = w_{T}(t) \sum_{k=1}^{52} s_k \exp(j2 \pi k {\Delta}_f t),
  \label{st}
\end{equation}
where ${\Delta}_f=312.5$~kHz is the frequency spacing between adjacent subcarriers and $s_k$ is the $k$-th element of $S$. Moreover, $w_{T}(t)$ is a window function that extends over multiple periods of FFT, and it is defined as
\begin{equation}
  w_{T}(t) = 
  \begin{cases}
  sin^2(\frac{\pi}{2}(0.5+{\frac{t}{T_{TR}}})), & \text{if}\ - {\frac{T_{TR}}{2}}<t<\frac{T_{TR}}{2};\\
  1, & \text{if}\ {\frac{T_{TR}}{2}}\leq t<T-\frac{T_{TR}}{2};\\
  sin^2(\frac{\pi}{2}(0.5-{\frac{t-T}{T_{TR}}})), & \text{if}\  {T-\frac{T_{TR}}{2}}\leq t< T+\frac{T_{TR}}{2},
  \end{cases}
\end{equation}
where $T_{TR}$, the transition time, is about 100~ns and $T = 10 \times 0.8 \mu s = 8 \mu s$.
In this way, the STS signal $s_S(t)$ is transmitted over the air.
Affected by the wireless channel, the received STS signal $\widetilde{s}_S(t)$ can be obtained as
\begin{equation}
  \widetilde{s}_S(t) = w_{T}(t) \sum_{k=1}^{52} s_k h_{c_k} \exp(j2 \pi k {\Delta}_f t).
\end{equation}
Then, with a  sampling rate of 20~Msps, 160 IQ points can be sampled from $\widetilde{s}_S(t)$ at the receiver.
Motivated by this relation, we take the processed CSI data $\widetilde{H}$ into Equation~\eqref{st} to generate corresponding time-domain signals as
\begin{equation}
  x_S(t) = w_{T}(t) \sum_{k=1}^{52} s_k \widetilde{h}_k \exp(j2 \pi k {\Delta}_f t).
\end{equation}
From the above equation, we can observe that the transmitter's RF fingerprints are embedded in $x_S(t)$. 

Similarly, we take the processed CSI data into LTS signals.
A long OFDM training symbol $L \in \mathbb{C}^{52}$ uses all 52 subcarriers for channel estimation and can be expressed as 
\begin{equation}
  \begin{aligned}
  L=& \left\{1,\ 1,\ -1,\ -1,\ 1,\ 1,\ -1,\ 1,\ -1,\ 1,\ 1,\ 1,\ 1,\ 1,\ 1,\ -1, \right.\\ 
  &\left. -1,\ 1,\ 1,\ -1,\ 1,\ -1,\ 1,\ 1,\ 1,\ 1,\ 1,\ -1,\ -1,\ 1,\ 1,\ -1, \right.\\ 
  &\left. 1,\ -1,\ 1,\ -1,\ -1,\ -1,\ -1,\ -1,\ 1,\ 1,\ -1,\ -1,\ 1,\ -1 \right.\\ 
  &\left. 1,\ -1,\ 1,\ 1,\ 1,\ 1\right\}.
    \end{aligned}
\end{equation}
The standard LTS signal $s_L(t)$ can be generated by
\begin{equation}
  s_L(t) = w_{T}(t) \sum_{k=1}^{52} l_k \exp(j2 \pi k {\Delta}_f (t-T_{GI2})),
  \label{lt}
\end{equation}
where $l_k$ is the $k$-th value of $L$. 
$T_{GI2}=1.6~\mu s$ is the training symbol guard interval duration.
Correspondingly, an LTS signal with RF fingerprints can be obtained by
\begin{equation}
  x_L(t) = w_{T}(t) \sum_{k=1}^{52} l_k \widetilde{h}_k \exp(j2 \pi k {\Delta}_f (t-T_{GI2})).
\end{equation}

Toward this end, we add $x_S(t)$ and $x_L(t)$ together and have the generated time-domain signal as
\begin{equation}
  x(t) = x_S(t)+x_L(t-T).
\end{equation}
Finally, by sampling it at a rate of 20~Msps, we obtain a time-domain feature vector $U\in \mathbb{C}^{320}$ as
\begin{equation}
      U = [u_1,\ u_2,\ \dots,\ u_{320}].
\end{equation}
In this way, we convert a CSI measurement of 52 subcarriers into a high-dimensional feature vector, carrying RF hardware imperfection characteristics. 

\subsection{Feature Extraction and Recognition}

\begin{figure}[t]
  \centering
  \includegraphics[width=\textwidth]{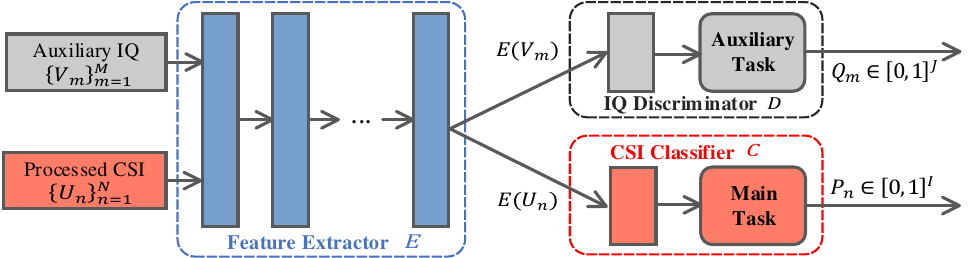}
  \caption{Workflow of the auxiliary learning model.}
  \label{fig:model_architecture}
\end{figure}

Although CSI measurements can be transformed into higher-dimensional feature vectors, 
they still suffer from losing fingerprint information incurred by processes like FFT.
To overcome this challenge, we exploit an auxiliary learning approach for CSI-based RF fingerprinting.
Specifically, we consider the task of fingerprinting registered devices as a multiclass classification problem $(\mathcal{U},\mathcal{D})$.
Therein, $\mathcal{U} = \left\{U_n\right\}^{N}_{n=1} \subseteq \mathbb{C}^{320} $ is the dataset of CSI feature vectors and can be extracted from the collected CSI measurements $\mathcal{H} =  \left\{H_n\right\}^{N}_{n=1} $.
In addition, we define an auxiliary IQ dataset $\mathcal{V} = \left\{V_m\right\}^{M}_{m=1} \subseteq \mathbb{C}^{320} $ that contains raw IQ samples from $J$ wireless devices $\mathcal{D}^v$.
$\mathcal{D}$ could be a subset of  $\mathcal{D}^v$, i.e., $\mathcal{D} \subseteq \mathcal{D}^v$, or they do not have common elements, i.e., $\mathcal{D} \cap \mathcal{D}^v = \varnothing$. 

For effective feature classification, we propose a dual-task learning model, as shown in Figure~\ref{fig:model_architecture}.
The proposed model consists of three components, including a feature extractor $E$, a CSI classifier $C$, and an IQ discriminator $D$. 
Specifically, the feature extractor $E$ is the front block of our model and provides shared layers for the both classifier and discriminator.
It takes a feature vector $U_n$ or an IQ sample $V_m$ as input and returns a hidden feature representation as $E(U_n)$ or $E(V_m)$.
The CSI classifier $C$ acts as the end block of the model and is specific to the main task of recognizing devices.
It takes a learned feature representation $E(U_n)$ as input and outputs an $I$-dimensional probability vector $P_n \in [0,1]^{I}$ in terms of all devices.
Similar to $C$, the IQ discriminator $D$ aims to complete the auxiliary task of identifying devices from $\mathcal{D}^v$, and it takes a hidden feature representation $E(V_m)$ as input and produces a $J$-dimensional probability vector $Q_m \in [0,1]^{J}$.

In the training phase, we adopt the cross-entropy as the loss function for both the classifier and discriminator.
%Generally, the cross-entropy calculates the average difference between predicted probabilities and ground-truth labels and has the following advantages.
%First, it penalizes incorrect predictions more heavily, forcing the model to focus on learning correct classes.
%Second, it provides a smooth and continuous optimization landscape, facilitating faster convergence during training.
%For the above reasons, the loss function of the main task $\mathcal{L}_{main}(E, C)$ is given as follows
The loss function of the main task $\mathcal{L}_{main}(E, C)$ is given by
\begin{equation}
  \mathcal{L}_{main} (E, C) = -{\frac{1}{N}} \sum_{n=1}^{N} \sum_{i=1}^{I} D^i_n \log(P^i_n),
\end{equation}
where $D^i_n$ denotes the binary indicator (0 or 1) of whether the $n$-th sample belongs to the $i$-th device, and $P^i_n$ represents the predicted probability of the $n$-th belonging to the $i$-th device. 
Similarly, the loss function of the auxiliary task  $\mathcal{L}_{auxi}(E, D)$  is given by 
\begin{equation}
  \mathcal{L}_{auxi}(E,D) = -{\frac{1}{M}} \sum_{m=1}^{M} \sum_{j=1}^{J} D^j_m \log(P^j_m).
\end{equation}

To improve the classification performance of our CSI classifier, it is crucial to implement an auxiliary learning strategy.
%The basic idea is that since CSI measurements and IQ samples are inherently correlated, the shared feature extractor can learn how to effectively extract RF fingerprints from raw IQ samples and transfer this knowledge to the main task when processing time-domain CSI feature vectors. 
To achieve this, a cooperative game is set between $E$, $C$, and $D$ during training.
Both $C$ and $D$ work together with $E$ to minimize their losses $\mathcal{L}_{main}$ and $\mathcal{L}_{auxi}$, respectively.
This cooperative training game enables the model to effectively utilize both types of data and learn more diverse and comprehensive device fingerprint features.
We implement the auxiliary learning approach  by solving the following minimization problem as
\begin{equation}
    \underset{E,C,D}{\min} \mathcal{L}_{main} (E, C) + \lambda \mathcal{L}_{auxi}(E,D),  
\end{equation}
where $\lambda>0$ is a hyperparameter.

%During the training phase, we initially optimize only the loss function of the main task for the first 50 epochs.
%In the subsequent 50 epochs, we optimize the two loss functions simultaneously.
%The reason for this is that by gradually introducing the auxiliary task, our model can focus on parameter optimization for the main task at the initial stage, which helps to avoid interference from the auxiliary task.
%This strategy allows better control over the optimization direction and priority of the whole model during the training phase.
%After enough training iterations, our model can achieve a more robust and accurate device classification performance when relying on CSI measurements alone.

\section{Experimental Evaluation}

\subsection{Evaluation Methodology}

\subsubsection{Auxiliary IQ Dataset.}
We take raw IQ samples from the WiSig dataset~\cite{11} to obtain our auxiliary dataset.
Specifically, the WiSig dataset consists of transmission signals from 174 commercial WiFi 802.11 devices captured by 41 USRPs.
%All devices are set to operate on channel 11 with a center frequency of 2462~MHz and a bandwidth of 20~MHz.
The USRPs capture WiFi signals at a sampling rate of 25~Msps in four different days within a month. 
In this condition, we choose 85 transmitters and 10 receivers and take their IQ samples from the WiSig dataset.
Then, we conduct energy detection, signal segmentation, and resampling to generate 320-point IQ samples.
For each transmitter-receiver pair, we select 30 samples, resulting in a total of 300 samples for each transmitter.
After that, our auxiliary IQ dataset contains more than 25K IQ samples.

\subsubsection{Synthetic CSI Dataset.}
Based on the auxiliary IQ dataset, we proceed to generate CSI measurements by performing the standard channel estimation on the corresponding IQ samples.
To do this, we perform channel estimation using the minimum mean square error method to obtain a dataset that involves about 25K CSI measurements for 85 wireless devices.

\begin{figure}[t]
  \centering
  \includegraphics[width=\linewidth]{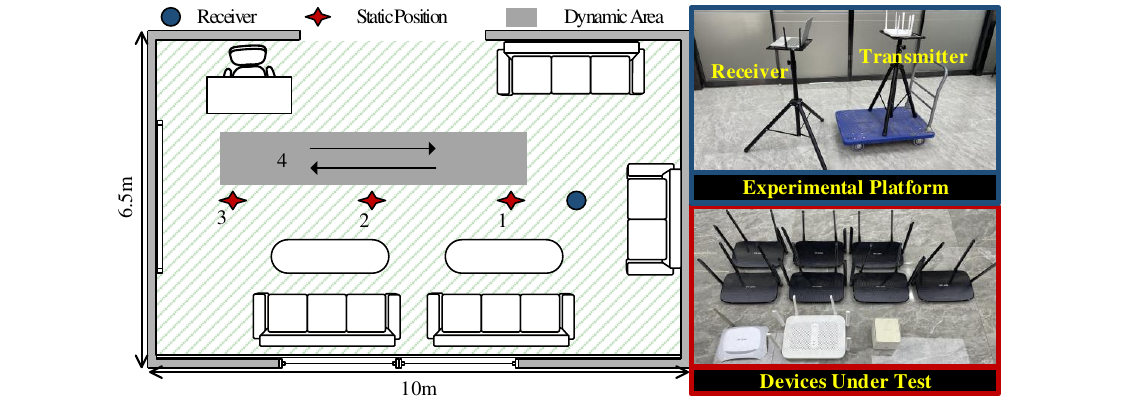}
  \caption{Experimental setting for CSI collection in a meeting room.}
  \label{fig:experimental_setting}
\end{figure}
\subsubsection{Real CSI Dataset.}
Besides the synthetic CSI dataset, we collect CSI measurements from ten off-the-shelf WiFi devices in a real-world environment. 
Figure~\ref{fig:experimental_setting} demonstrates the experimental setting for CSI collection.
Specifically, the devices under test include one Honor router, one Redmi router, and eight TP-Link routers.
The receiver is a laptop equipped with an Intel WiFi 6E AX211 network interface card.
The PicoScenes~\cite{31}, a CSI collection tool, is running on the laptop to capture WiFi frames from all routers.
We mount the receiver and transmitters with a height of 1.2~m in a meeting room with a size of 6.5~m $\times$ 10~m.
When collecting CSI measurements, we first place each transmitter 1~m, 3~m, and 5~m away from the receiver, respectively.
Then, we keep the transmitter in a moving state.
In each scenario, we collect 50 CSI measurements from each router, resulting in a total of 200 samples per device.
Finally, we obtain a real CSI dataset with 2K samples.

\subsubsection{Implementation.}
After obtaining datasets, we implement the proposed system in a desktop computer that is equipped with an Intel(R) Core(TM) i5-9400 processor and runs Windows 11.
We perform the process of CIM and TDSG in MATLAB R2016b, while the training and testing of dual-task models is conducted in PyTorch version 1.13.1, using Python version 3.7. 

\subsubsection{Evaluation Metrics.} We use the following metrics to evaluate the performance of our system. 
\begin{itemize}
    \item \textbf{Accuracy.} It is defined as the ratio of the number of samples that are correctly classified to the total number of samples.
    \item \textbf{F1 Score.} It is the harmonic mean of precision and recall, which comprehensively reflects the model's classification performance.
\end{itemize}

\subsection{Experimental Results}
\subsubsection{Evaluation on Synthetic CSI Dataset}
First, we demonstrate the difference between IQ and CSI data in RF fingerprinting.
For this purpose, we select six candidate models including convolutional neural network (CNN), recurrent neural network (RNN), temporal convolutional network (TCN), support vector machine (SVM), K-nearest neighbors (KNN), and random forest.
CNN and RNN are two mainstreaming deep learning networks in current RF fingerprinting solutions~\cite{8,20,33,34,35}.
TCN is a neural network that is good at processing sequential data. 
In addition, SVM, KNN, and random forest are widely used machine learning algorithms. 
In our exper-iment, both the CNN and TCN are configured with four convolutional layers, while the RNN has four recurrent layers. 
We directly train six candidate models on the aux-iliary IQ dataset and the synthetic CSI dataset. 
For each device, we randomly select 240 samples for training and the remaining 60 samples for testing. 
The initial learning rate is set to 0.001, and the epoch size is 100. 
The cosine annealing algorithm is uti-lized to gradually reduce the learning rate.

\begin{figure}[t]
  \centering
  \includegraphics[width=\textwidth]{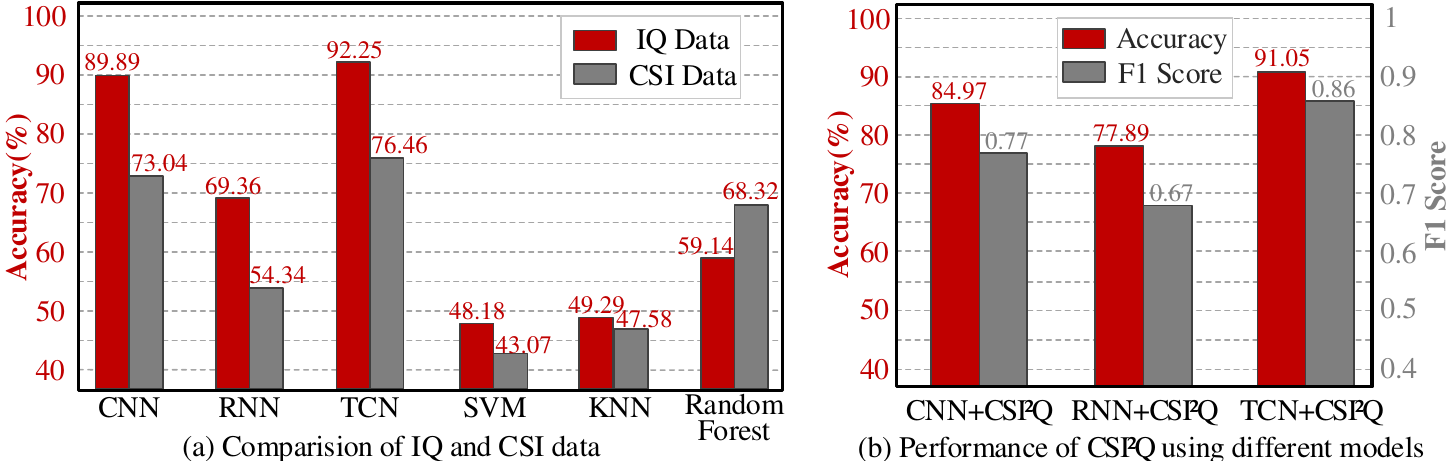}
  \caption{Classification results of different models.}
  \label{fig:model_comparison}
\end{figure}

Figure~\ref{fig:model_comparison}~(a) shows the classification performance of IQ and CSI data on different models. 
We can observe that, except for the random forest, the candidate models show a higher accuracy on IQ data than CSI data. 
The random forest produces contrasting results. 
This phenomenon can be attributed to the increasing sparsity of data points in higher-dimensional spaces as the feature dimensions grow. 
In such high-dimensional realms, the distances between individual samples tend to inflate, posing a challenge for the random forest classifier to delineate an effective decision boundary. 
Deep learning models achieve much better classification performance in RF fingerprinting than machine learning approaches. 
Because they can automatically extract high-level features from raw data, but machine learning approaches depend on handcrafted features and are unsuitable to recognize devices based on raw IQ and CSI samples.

Next, we present the performance of the proposed system using different deep learning models.
We build three feature extractors using four-layer CNN, RNN and TCN, respectively, each of which connects with a CSI classifier and an IQ discriminator. 
Each deep learning model has two fully connected layers. 
We take three-quarters of raw IQ samples and synthetic CSI measurements for training and the remaining for testing.
Figure~\ref{fig:model_comparison}~(b) illustrates the performance of CSI\textsuperscript{2}Q on three different deep learning models.
With the help of CSI\textsuperscript{2}Q, all models' classification performance is improved.
Among them, the TCN achieves the best performance, and its accuracy increases from 76\% to 91\%, which is comparable to that of the IQ-based TCN reported in Figure~\ref{fig:model_comparison}~(a).
The result demonstrates the effectiveness of our system in improving CSI fingerprinting.
Since the TCN outperforms the other models, the following experiments are conducted based on the TCN.

%\begin{table}
%    \caption{Performance of CSI\textsuperscript{2}Q on the synthetic dataset in the open-world scenario}
%    \label{tab:Open-world_recognition_Wisig} 
%    \begin{center}
%         \begin{tabular}{ | m{5.5cm} | m{3cm} | m{3cm} | }
%      \hline
%       Models & Accuracy & F1 Score\\
%      \hline
%       CSI & 71.13\% & 0.59\\
%       CSI + CSI\textsuperscript{2}Q & 87.28\% & 0.82\\
%       CSI + CSI\textsuperscript{2}Q (no OpenMax) & 83.79\% &0.77\\
%      \hline
%    \end{tabular}
%    \end{center}
%\end{table}

Lastly, we perform an ablation study to measure the significance of each component in our system.
We divide CSI\textsuperscript{2}Q into three parts, i.e., CIM, TDSG, and ALIQ.
We remove the three components one by one, train the remaining parts, and compare their influence on the system's classification accuracy and F1 score.
\begin{table}[t]
    \caption{Ablation study for CSI\textsuperscript{2}Q}
    \label{tab:tcn_ablation} 
         \begin{tabular}{ |c|c|c|c|c| }
      \hline
       \ \ \ \ \ \textbf{Method}\ \ \ \ \ \ &\ \ \ \ \ \  \textbf{CSI\textsuperscript{2}Q}\ \ \ \ \ \ \  &\  \textbf{TDSG + ALIQ}\ \  &\ \ \ \ \ \ \textbf{ALIQ}\ \ \ \ \ \ \  &\ \ \ \ \ \  \textbf{CSI}\ \ \ \ \ \ \ \\
      \hline
       Accuracy & 91.05\% & 88.23\% & 83.17\% & 76.32\%\\
       \hline
       F1 Score & 0.86 & 0.82 & 0.74 & 0.65\\
      \hline
    \end{tabular}
\end{table}
As shown in Table~\ref{tab:tcn_ablation}, each component has a positive impact on the overall performance.
The combination of CIM, TDSG, and ALIQ achieves a classification accuracy of 91\%. 
When removing CIM, the accuracy decreases to 88\%. When only ALIQ exist-ing, the accuracy decreases to 83\%. 
When removing all components, the accuracy drops to 76\%. 
A similar trend can be also observed in F1 score. 
The results of ablation experiments demonstrate that all system designs have a significant impact on the overall performance, particularly the TDSG and ALIQ.
Moreover, these experimental findings further affirm the effectiveness of CSI\textsuperscript{2}Q in improving CSI fingerprinting.

%In addition, we demonstrate the impact of the CSI measurement transformation on CSI measurements.
%Figure~\ref{fig:ablation_feature_visualization} illustrates the high-dimensional feature visualization of CSI data after applying our TDSG and CIM+TDSG methods.
%We can observe that the TDSG scheme reduces the intra-class distances and alleviates the overlap of inter-class feature points.
%Furthermore, when both the CIM and TDSG designs are applied, the features show a great reduction in intra-class distances and a significant increase in inter-class distances.
%These observations suggest that the time-domain representation of CSI samples is very helpful for CSI fingerprinting. 
%\begin{figure}[t]
%  \centering
%  \includegraphics[width=\textwidth]{abltaion_feature_visualization_0331.pdf}
%  \includegraphics[width=\textwidth]{feature_visualization_0424.pdf}
%  \caption{Feature visualization for CSI measurement transformation.}
%  \label{fig:ablation_feature_visualization}
%\end{figure}
\subsubsection{Evaluation on Real CSI Dataset}
\begin{figure}[t]
  \centering
  \includegraphics[width=\linewidth]{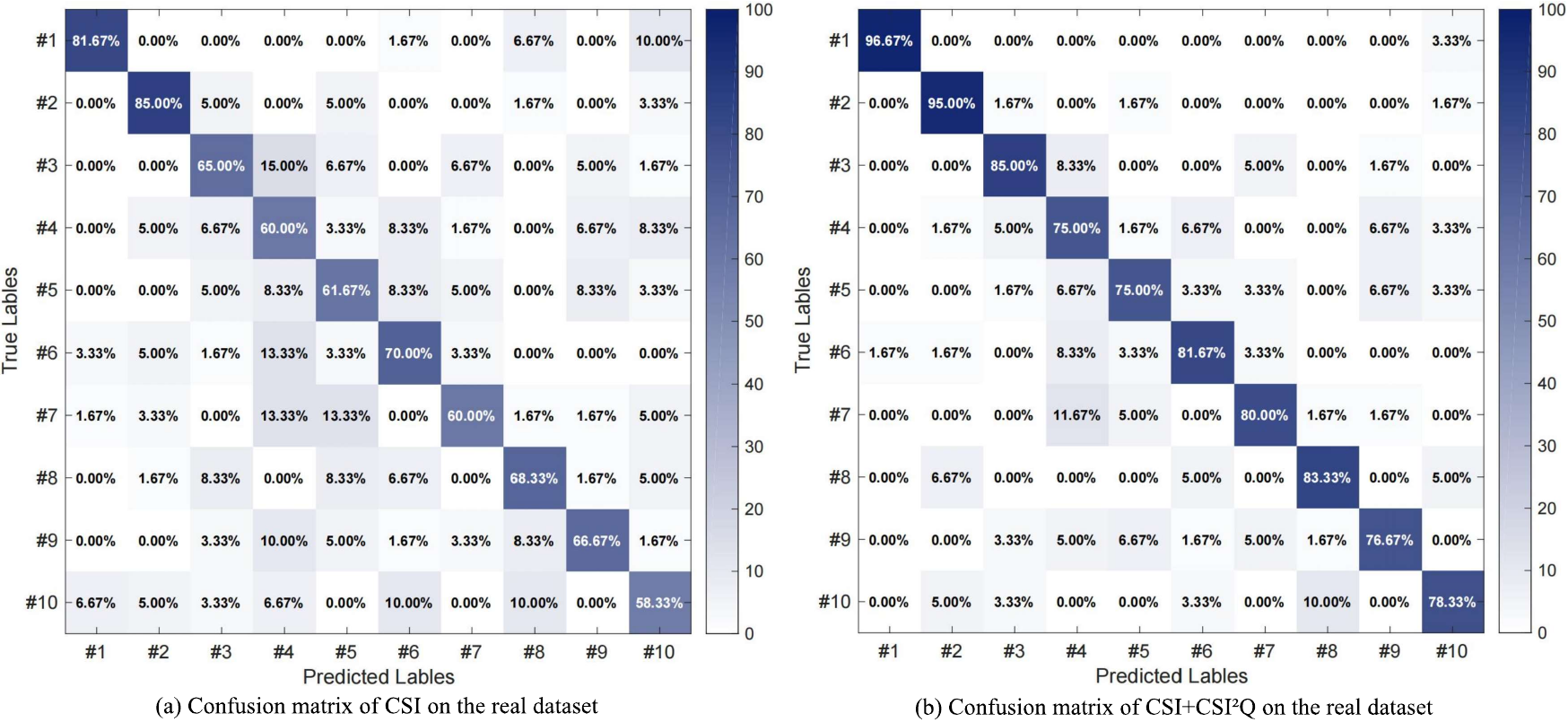}
  \caption{Confusion matrix of classification results on the real dataset. }
  \label{fig:confusion_matrix}
\end{figure}
The data in the synthetic CSI dataset only comes from static transmitters, while our CSI data includes measurements from moving transmitters. We still use the auxiliary IQ dataset from the WiSig. The training set comprises 1400 CSI samples from 10 devices, while the testing set consists of 600 samples.
Figure~\ref{fig:confusion_matrix} illustrates the impact of CSI\textsuperscript{2}Q on an individual transmitter.
The first two represent the Honor and Redmi routers and the remaining are TP-Link routers.
We can find that CSI\textsuperscript{2}Q system has significantly improved the classification accuracy of all devices, which can achieve an increase of up to 20\% and an average increase from 67\% to 82\%.
In addition, the Honor and Redmi devices exhibit the highest accuracy. 
The result also indicates there are significant differences among RF imperfections of different brands of WiFi devices.
As a result, the RF fingerprints of Honor and Redmi routers are more different from TP-Link devices.

\section{Conclusions}
This paper presents CSI\textsuperscript{2}Q to achieve comparable performance to IQ-based ap-proaches. 
Different from the existing CSI fingerprinting schemes that only focus on alleviating channel interference or extracting channel-independent fingerprint fea-tures from CSI measurements, we have solved channel interference, coarse granulari-ty and information loss three challenges at the same time. 
Instead of extracting RF fingerprints directly from raw CSI measurements, CSI\textsuperscript{2}Q opts for a transformative method. 
It converts the CSI data into time-domain signals, thereby aligning them with the feature space inherent to IQ samples. 
And the strong feature extraction abil-ity of an IQ fingerprinting model is transferred to its CSI counterpart via an auxiliary training strategy. 
Finally, the trained CSI fingerprinting model is used to decide which device the sample under test comes from. 
Experimental results show that our system can effectively improve the performance of CSI fingerprinting with an accuracy improvement of about 15\%.

\bibliographystyle{splncs04}
\bibliography{references}
\end{document}